\documentclass{rnaastex}
\usepackage{hyperref}

\newcommand{\TableURL}{https://zenodo.org/records/17360103?token=eyJhbGciOiJIUzUxMiJ9.eyJpZCI6ImYzNzNkOTI3LTQzZDYtNDBlZS1iODI2LTY1NDUwOGI3ZjcxOSIsImRhdGEiOnt9LCJyYW5kb20iOiJmMTc0MDJhOGQ4ZjE5NzkwMjU4YmUzODM5ZmNiZmFkMiJ9.LYc2mQ5NvTKO3QgWdCvlKZuSmrX7wsRNMZhkGXF4TBfk1r5cl1ymKFQLPUO11ryv5VYmEkySagf_FacxYtfX4w}
\newcommand{\LetterURL}{https://zenodo.org/records/17360286?token=eyJhbGciOiJIUzUxMiJ9.eyJpZCI6ImZlMGViZTc0LTgxZWMtNDRhMC1hMmI3LWYxNmY2Zjc0MDcxZCIsImRhdGEiOnt9LCJyYW5kb20iOiI4MTFiMzk3NTAzZDQ0YWUyY2U4Y2E2NGNiMzk3YjQyOCJ9.BCEX4sUDGO2K9t2wDIL2mk9iVyRe7f19lJGFyau0nabe-lxKGMsrUTaq0WIIYZysWFIHoUhRWdTwxVf9CZQQSQ}
\newcommand{\RNURL}
{http://iopscience.iop.org/article/10.3847/2515-5172/aaa12e}

\begin{document}

\title{The Third Workshop on Extremely Precise Radial Velocities: The New Instruments}

\correspondingauthor{Jason T.\ Wright}
\email{astrowright@gmail.com}
 
\author[0000-0001-6160-5888]{Jason T.\ Wright}
\affil{Department of Astronomy \& Astrophysics / 525 Davey Laboratory /
The Pennsylvania State University /
University Park, PA, 16802, USA}
\affil{Center for Exoplanets and Habitable Worlds / 525 Davey Laboratory /
The Pennsylvania State University /
University Park, PA, 16802, USA}

\author[0000-0003-0149-9678]{Paul Robertson}
\affil{Department of Astronomy \& Astrophysics / 525 Davey Laboratory /
The Pennsylvania State University /
University Park, PA, 16802, USA}
\affil{Center for Exoplanets and Habitable Worlds / 525 Davey Laboratory /
The Pennsylvania State University /
University Park, PA, 16802, USA}
\affil{NASA Sagan Fellow}


\section{The Third Workshop on Extremely Precise Radial Velocities} 

The Third Workshop on Extremely Precise Radial Velocities was held at the Penn Stater Conference Center and Hotel in State College, Pennsylvania, USA from 14--17 August 2016, and featured over 120 registrants from around the world. The talks were live-streamed to facilitate remote participation, and slides from the plenary sessions will be available at the conference website.\footnote{\url{http://eprv2017.psu.edu/}}. The SOC chair was Jason T.\ Wright; the LOC chair was Paul Robertson.

The talk follows two prior successful workshops. The first was Astronomy of Exoplanets with Precise Radial Velocities, held from 16--19 August 2010 also in State College, which included a historical talk by Gordon Walker, a discussion on the future of precise RVs in the USA, and the generation of a letter from the community to NASA and the NSF in response to the 2010 decadal survey.\footnote{\href{\LetterURL}{At this URL.}} The Second Workshop on Extreme Precision Radial Velocities was held in New Haven, Connecticut, USA from 5--8 July 2015, and the discussions there on the state of the art of the field resulted in a white paper \citep{Fischer}. Like its two predecessors, the third workshop sought to promote a frank exchange of technical ideas on how to improve the detection of exoplanets via precise radial velocities. The motto of these workshops has been (to mix metaphors) ``nuts and bolts, warts and all.'' At the valediction of the this installment, Xavier Dumusque and the Geneva exoplanet group announced plans to hold a fourth workshop in early 2019 in Switzerland.

Highlights of the third workshop included an introductory historical talk by Paul Butler, ``poster pops'' for 15 minutes on the first two days where participants briefly discussed their contributed posters, and a wrapup of the plenary sessions by Peter Plavchan and Cullen Blake.

The conference included five persistent breakout sessions that met concurrently during one 90 minute block on three of the four days. These sessions were ``mini-conferences'' organized by their chairs to be highly technical, ``deep dives'' into the current challenges and solutions to problems that limit the state of the art.  The sessions covered: Hardware Challenges and Comprehensive Error Budgets (chairs Sam Halverson and Arpita Roy), Stellar Photospheres and Modeling Jitter (chair Heather Cegla), Statistical Methods (chair Rodrigo Diaz), Computational Methods (chairs Eric Ford and Benjamin Nelson), and Observational Strategies (chair Jennifer Burt). The session chairs briefly presented discussion summaries in the plenary session at the end of each day they met.

\section{The New Instruments}

The second day was dedicated entirely to brief, technical presentations by representatives from 23 instrument teams building or commissioning new instruments. In addition to their 10 minute talks, each representative was invited to populate a spreadsheet with a summary of their instrument's important characteristics.  We present a the complete spreadsheet in Figure 1 below and as a machine-readable table in our Research Note \href{\RNURL}{here}.\footnote{A standalone PDF of the table is available at \href{\TableURL}{this URL}.} This table describes nearly all\footnote{No representatives of the HRS2/HET, PEPSI/LBT, or HIRES/ELT teams could attend.} of the next-generation extremely precise Doppler velocimeters being designed, built, or commissioned today, and complements an earlier table compiled by \citet{Pepe}

\acknowledgments

The workshop organizers acknowledge the valuable financial and logistical support of its sponsors: Isotech, NEOS, Princeton Infrared Technologies, Semiconductor Technology Associates, Inc., SAMSI, NExSI, the NSF (Solar and Planetary Research Grant 1734032), NASA (Planetary Science Division grant NNX17AK77G), the Penn State Center for Exoplanets and Habitable Worlds, and the Penn State Institute for Cyber Science. 
\begin{figure}
\caption{List of new instruments. This table is available in machine readable format in our \href{\RNURL}{Research Note} and a standalone PDF is \href{\TableURL}{here}.}
\includegraphics[angle=90,width=7.5in]{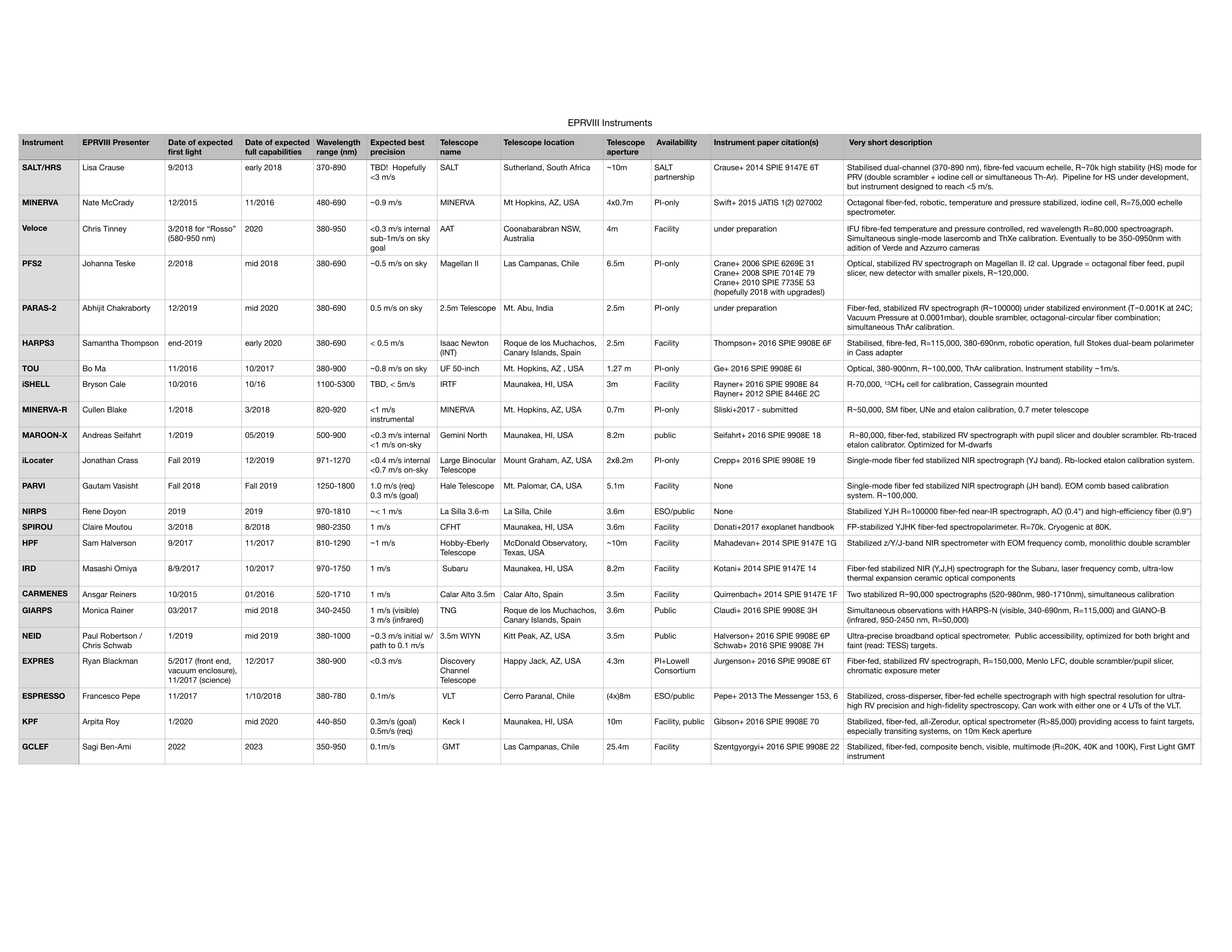}
\end{figure}


\begin{thebibliography}{}
\bibitem[Fischer et al.(2016)]{Fischer} Fischer, D.~A., Anglada-Escude, G., Arriagada, P., et al.\ 2016, \pasp, 128, 066001 

\bibitem[Pepe et al.(2014)]{Pepe} Pepe, F., Ehrenreich, D., \& Meyer, M.~R.\ 2014, \nat, 513, 358 


\end{thebibliography}
\end{document}